\documentclass[a4paper,11pt]{article}
\usepackage{jinstpub}

\usepackage{graphicx}
\usepackage{siunitx}

\usepackage[european]{circuitikz}
\usetikzlibrary{calc}
\usetikzlibrary{shapes}
\usetikzlibrary{backgrounds}
\usetikzlibrary{automata,positioning}

\DeclareSIUnit\bit{b}
\DeclareSIUnit\Byte{B}
\DeclareSIUnit\sample{S}
\DeclareSIUnit\lsb{lsb}
\DeclareSIUnit\cps{cps}

\usepackage{lineno}

\title{The Remote Analog to Digital Conversion DAQ System for the TRISTAN Detector Upgrade}

\author[a,b,1]{A.S. Gavin,\note{Present address: Max-Planck-Institut für Kernphysik, Saupfercheckweg 1, 69117 Heidelberg, Germany}}
\author[c]{M. Balzer,}
\author[c]{S. Chilingaryan,}
\author[a,b]{R. Henning,}
\author[c]{A. Kopmann,}
\author[d]{S. Mertens,}
\author[c]{J. Mostafa,}
\author[c]{F. Simon,}
\author[c]{N. Tan Jerome,}
\author[c]{D. Tcherniakhovski,}
\author[e,2]{K. Urban,\note{Present address: Politecnico di Milano, Dipartimento di Elettronica, Informazione e Bioingegneria, Piazza L. da Vinci 32, 20133 Milano, Italy}}
\author[a,b]{J. F. Wilkerson,}
\author[c]{S. Wüstling}

\affiliation[a]{Department of Physics and Astronomy, University of North Carolina, Chapel Hill, NC 27599, U.S.A.}
\affiliation[b]{Triangle Universities Nuclear Laboratory, Durham, NC 27708, U.S.A.}
\affiliation[c]{Institute for Data Processing and Electronics (IPE), Karlsruhe Institute of Technology (KIT), Hermann-von-Helmholtz-Platz 1, 76344 Eggenstein-Leopoldshafen, Germany}
\affiliation[d]{Max-Planck-Institut für Kernphysik, Saupfercheckweg 1, 69117 Heidelberg, Germany}
\affiliation[e]{Department of Physics, TUM School of Natural Sciences, Technical University of Munich, James-Franck-Straße 1, 85748 Garching b. München, Germany }

\emailAdd{andrew.gavin@mpi-hd.mpg.de}

\abstract{The TRISTAN detector is an upgrade to the KATRIN experiment to enable a differential measurement of the tritium $\upbeta$-decay spectrum to search for sterile neutrinos with $\unit{\kilo\eV}$ masses.
This entails performing precision electron spectroscopy with over one thousand silicon drift detector pixels, each responsible for recording incident electron rates of $10^5$ counts per second. 
A project-specific data acquisition (DAQ) system is developed to meet the experimental challenges through a remote analog to digital conversion (RADC) design. 
In this work, the conceptual design of the RADC DAQ is presented along with the built system for operating the TRISTAN detector upgrade.
The system includes flexible signal processing logic and data management that is optimized for the high-rate precision measurement. 
}


\arxivnumber{}

\begin{document}

\maketitle
\flushbottom

\section{Sterile Neutrino Search with the TRISTAN Detector Upgrade}\label{sec:introduction}

A natural extension to the Standard Model is the addition of right-handed sterile counterparts to the observed left-handed neutrinos \cite{Abazajian:2012ys}.
Light sterile neutrinos, with $\SI{}{\electronvolt}$ scale masses, have been proposed as partial solutions to existing experimental neutrino anomalies, while heavier $\SI{}{\kilo\electronvolt}$ mass sterile neutrinos serve as a viable candidate for part of the dark matter \cite{Adhikari_2017,Boyarsky_2019,Acero_2024}. 
A precision measurement of the $\upbeta$-decay spectrum can be used to constrain the mixing of active to sterile neutrinos through their distinctive kink-like signature.

The KATRIN experiment measures the endpoint region of the tritium $\upbeta$-decay spectrum to place leading limits on the effective active neutrino mass \cite{KATRIN:2021uub,KATRIN_2025}. 
A search for shape distortions in the measured spectrum have been used to constrain eV scale sterile neutrinos, while additional measurements were used to expand the search up to $\approx \SI{1}{\kilo\electronvolt}$ \cite{Acharya2025,Aker2023_keV}. 
Constraining higher mass sterile neutrinos to lower mixing requires measuring a wider energy range of the spectrum and better statistical scaling with measurement time.
The TRISTAN detector upgrade to the KATRIN experiment is designed to perform a high precision differential measurement of the full tritium spectrum, with a goal of improving the statistical sensitivity to $\SI{}{\kilo\electronvolt}$ mass sterile neutrinos to a mixing of $\sin^2 (\theta) < 10^{-6}$ ($95\%$ C.L.) \cite{Mertens_2019}.

A highly pixelated silicon drift detector (SDD) array was chosen for this application and built at the Semiconductor Laboratory (HLL) of the Max Planck Society \cite{Lechner2004_SDD}.
A TRISTAN detector module is composed of a monolithic array of 166 hexagonally tiled $\SI{7}{\milli\meter^2}$ pixels.
Each pixel is designed to accurately reconstruct the energy of $\SI{20}{\kilo\electronvolt}$ electrons with a resolution $\SI{300}{\electronvolt}$ while handling incident electron rates of up to $10^5$ counts per second.
An installation of nine modules, totaling 1494 pixels, in the KATRIN detector section is planned for the $\SI{}{\kilo\eV}$ sterile neutrino search \cite{Mertens_2019}.
The operation of the TRISTAN detector must ensure the ability to perform the high rate and high density readout of channels with minimal distortion to the reconstructed energy of electrons measured in the detector.
In this work a novel remote analog to digital conversion (RADC) data acquisition (DAQ) system is presented as a robust and scalable operation and readout architecture for the TRISTAN detector upgrade.

\section{Event Energy Reconstruction}\label{sec:filters}

The sensitivity of the TRISTAN detector upgrade hinges on the ability to perform high precision electron spectroscopy at high rates with the SDDs. 
This requires careful considerations in not only the detector design but also the data acquisition chain and pulse processing treatment.

A TRISTAN detector module consists of a monolithic SDD chip attached perpendicularly to a copper cooling and support structure, which hosts the in-vacuum electronic readout components \cite{Siegmann_2024}.
Signal amplification and detector control are provided by the ETTORE ASIC, which is a 12-channel low-noise charge sensitive amplifier with a pulsed reset tailored to operate the TRISTAN SDD with an integrated junction field-effect transistor \cite{ETTORE_ASIC}.
Energy depositions in the TRISTAN detector from incident electrons create electric pulses which have heights proportional to the deposited energy.
The pulses consist of a a fast risetime, $\tau_{\text{rise}}\sim \SI{100}{\nano\second}$, and subsequent decaying tail, with a decay constant of $\tau_{\text{decay}}\sim \SI{15}{\micro\second}$.

\begin{figure}[ht!]
\includegraphics[width=\columnwidth]{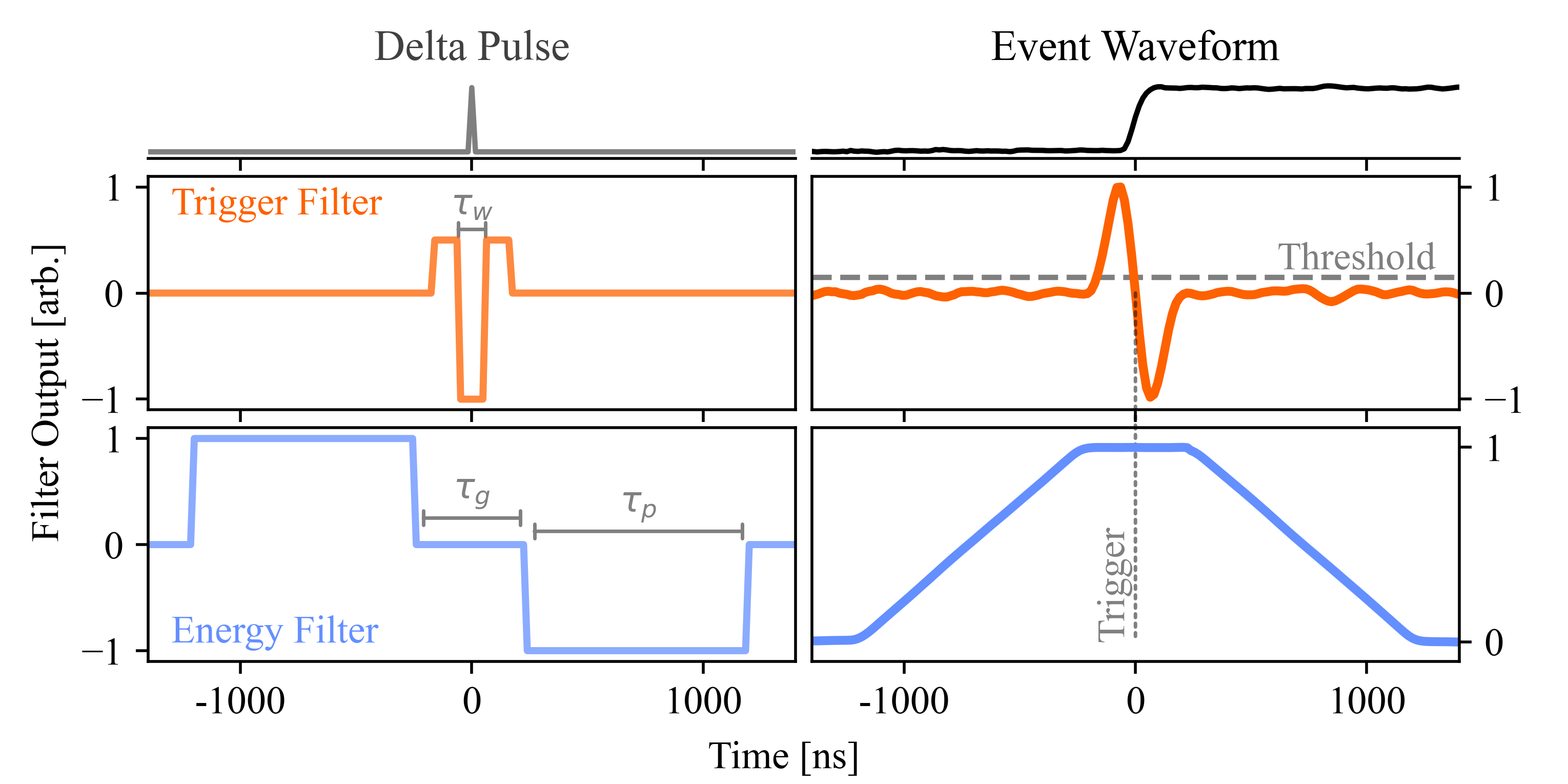}
\caption{Example response of the trigger (top) and energy (bottom) filters to delta pulse and detector event waveforms. The filters are parameterized by widths $\tau_w$, $\tau_g$, and $\tau_p$. An event trigger is produced and the energy filter is evaluated at the first zero crossing of the trigger filter after a threshold crossing.}
\label{fig:filter_performance}
\end{figure}

A full waveform digitization and subsequent application of finite impulse response (FIR) filters is necessary to accurately reconstruct the timing and height of the pulses \cite{Dolde_2017}. 
A combination of two FIR filters, a fast trigger filter and an energy filter, are used to have optimized performance for event pileup rejection and energy reconstruction \cite{Knoll:1300754}.
The response of the trigger and energy filter to a delta impulse and a detector event waveform are displayed in \autoref{fig:filter_performance}.
In order to have multiple samplings over the rising edge of the pulse, a digitization rate of more than $\SI{10}{\mega\hertz}$ is necessary. 

A fast bipolar filter, with time width $\tau_w$, is used to trigger the readout of the detector and distinguish between coincident events.
The response to an event occurring within the detector pixel is a heartbeat-like signal with an amplitude proportional to the pulse height.
The trigger logic functions by setting the filter to an armed state after crossing a fixed threshold, and producing a trigger timestamp at the next zero crossing.
Unlike a pure threshold trigger applied to a trapezoidal filter, this method benefits from timestamp evaluations that are independent of event energy.

The response length of the trigger filter is proportional to $\tau_w$ and therefore a minimization of this length is useful to separate the response between coincident detector events.
However, $\tau_w$ must remain longer than the detector risetime $\tau_{\text{rise}}$ to produce the necessary heartbeat response.
The setpoint for the trigger threshold is determined by the tradeoff between acceptance of low energy electrons and the introduction of random noise triggers.
With the measured TRISTAN detector noise performance and $\tau_w = \SI{112}{\nano\second}$, a trigger threshold set to accept $99\%$ of $\SI{2}{\kilo\electronvolt}$ energy depositions corresponds to an $8.5 \sigma$ fluctuation of the readout noise \cite{Urban_dissertation}.
The threshold value can be tuned for individual measurements given their measurement goals and sensitivity to noise triggers.

The deposited energy of a triggered event is reconstructed using pulse height analysis consisting of an exponential deconvolution and a trapezoidal filter.
A digital offset is subtracted from the waveform to correct for the detector working point voltage and analog offset applied before digitization.
An exponential deconvolution is then applied with a decay constant of $\tau_{d}$, which produces a more stable step-like pulse for the energy evaluation.

The energy filter is a trapezoidal filter defined by its shaping time, $\tau_p$, and gap time, $\tau_g$, parameters.
Larger values of the filter widths provide better energy resolution and a more stable filter response but lengthen the filter response time. 
For the TRISTAN detector operation values of $\tau_p = \SI{960}{\nano\second}$ and $\tau_g = \SI{480}{\nano\second}$ are set to match the goal energy resolution and mitigate pileup effects.

If two electrons deposit energy in a pixel within the response time of the energy filter a bias in both energy evaluations is introduced due to the overlap of the trapezoidal responses.
At increasing rates or increasing the filter response times the fraction of events that contribute to this pileup spectrum increases.
For a TRISTAN detector operating with $10^5$ electrons incident per pixel per second, pileup rates of $\mathcal{O}(10\%)$ are expected and therefore must be accurately flagged by the filter logic.

When two triggers occur within a set pileup window, $w_{\text{pu}}$, the events are flagged as pileup.
These events can be excluded from the analysis to remove the evaluation bias and are considered resolved pileup events. 
Coincident events occurring within the risetime of the detector however produce a variable number of triggers depending on the event energies and time separations.
To avoid biasing the spectrum, a trigger holdoff, $w_{\text{h}} \approx \tau_w$, is applied, which inhibits the creation of multiple events within the set time window.
These events are indistinguishable from single energy depositions in the filtering algorithm and remain in the measured spectrum as unresolved pileup.
The gap time of the energy filter is then tuned to ensure that the energy evaluation of such events corresponds to the sum of all energy depositions within the holdoff period.

In addition to coincident events within the same pixel, inter-pixel effects require trigger information to be shared between pixels.
A map of physically neighboring pixels allows the flagging of coincident events biased by charge sharing that occurs from electrons incident near the pixel borders.
Events that are biased by other inter-pixel effects, such as crosstalk between readout lines, can also be flagged by dedicated maps built from characterization measurements. 
Events affected by detector saturation, detector reset periods, or other waveform distortions are also flagged.

\section{Remote Analog to Digital Conversion Concept}\label{sec:RADC}

The DAQ of the TRISTAN detector modules must be a robust system to provide the low distortion digitization and event filtering necessary to operate a large number of pixels at high event rates. 
The structure is based on early signal digitization by front-end boards and subsequent signal processing on back-end field-programmable gate array (FPGA) boards. 
The design concept of this custom remote analog to digital conversion (RADC) DAQ is shown in \autoref{fig:RADC_concept}.

\begin{figure}[h!]
\centering
\tikzstyle{every node}=[font=\normalsize]
\tikzstyle{bag} = [align=center]

\tikzstyle{DETECTOR} = [draw, fill= violet!20 , rectangle, minimum height=3.5cm, minimum width=3.5cm]
\tikzstyle{SDD} = [draw, fill=blue!20, rectangle, minimum height=2cm, minimum width=1.5cm]
\tikzstyle{ASIC} = [draw, fill=blue!10, rectangle, minimum height=2cm, minimum width =1.5]

\tikzstyle{Front-end} = [draw, fill=red!20, rectangle, minimum height=3.5cm, minimum width=5cm]
\tikzstyle{ADC} = [draw, fill=orange!20, rectangle, minimum height=1.25cm, minimum width=2cm]
\tikzstyle{SubModule} = [draw, fill=orange!20, rectangle, minimum height=0.50cm, minimum width=2cm]

\tikzstyle{FPGA} = [draw, fill=yellow!20, rectangle, minimum height=1.5cm, minimum width=2cm]

\tikzstyle{Back-end} = [draw, fill=teal!20, rectangle, minimum height=3.5cm, minimum width=2cm]
\tikzstyle{Card} = [draw, fill=green!15, rectangle, minimum height=1.5cm, minimum width=1cm]

\tikzstyle{Server} = [draw, fill=green!25, rectangle, minimum height=0.50cm, minimum width=1cm]

\tikzstyle{VFT} = [draw, fill=black!20, rectangle,rotate=90,minimum height=0.5cm,minimum width=3.5cm,text=black!75]

\resizebox{0.95\textwidth}{0.3\textwidth}{
\begin{tikzpicture}[auto,node distance=0.1 cm]
    \node (detector) at (0,0) [DETECTOR] {};
    \node (sdd) at ($(detector.west)+(1.15,0.0)$) [SDD] {SDD};
    \node (asic) at ($(sdd.east)+(0.60,0.0)$) [ASIC] {ASICs};
    \node [below,font=\bfseries] at (detector.north){TRISTAN Detector};
    \node (tmb)      at (6,0) [Front-end] {};
    \node[below,font=\bfseries] at (tmb.north){Front-end};
    \node (adcs)     at ($(tmb)+(-1.3, 0.6)$) [ADC] {ADCs};
    \node (supply)   at ($(tmb)+(-1.3,-0.90)$) [SubModule] {Bias supply};
    \node (fpga)     at ($(tmb)+( 1.3, 0.25)$) [FPGA] {FPGA1};
    \node (muC)      at ($(tmb)+( 1.3, -1.25)$) [SubModule] {$\mu C$ System};

\foreach \a in {2,...,5}
    \draw[-Stealth] ($(asic.east)+(0,+\a*0.12)$) -- ++ ($(1,0) + (-\a*0.12,0)$) |- ($(adcs.west)+(0,\a*0.12-0.42)$);
\foreach \a in {2,...,5}  
    \draw[-Stealth] ($(adcs.east)+(0,-0.42+\a*0.12)$) -- ++ ($(-0.04,0) + (\a*0.08,0)$) |- ($(fpga.west)+(0,\a*0.24-0.84)$);

\foreach \a in {2,...,3}
    \draw[Stealth-] ($(asic.east)+(0,-\a*0.1)$) -- ++ ($(1,0) + (-\a*0.1,0)$) |- ($(supply.west)-(0,\a*0.1-0.25)$);

\foreach \a in {4}
    \draw[-Stealth] ($(asic.east)+(0,-\a*0.1)$) -- ++ ($(1,0) + (-\a*0.1,0)$) |- ($(muC.west)-(0,\a*0.1-0.3)$);   

\draw [-Stealth] ($(muC.west) + (0,0.1)$) --++ (-0.3,0) |- (supply.east);

    \node (vft)      at (3,0.0) [VFT] {Vacuum Feedthrough};
    \node (crate)      at (12,0) [Back-end] {};
    \node[below,font=\bfseries] at (crate.north){Back-end};
    \node (card)      at ($(crate)+(0,0.25)$)[Card]  {FPGA2};
    \node (server)      at (12,-1.25) [Server]  {Server};

    \node at (9.75,0.25) [text width=1cm,align=center,text=black!75] {optical links};

\draw[Stealth-Stealth] (muC.east) -- (server.west);   
\draw[Stealth-Stealth] (card.south) -- (server.north);   

    \node at (9.75,-1.45) [text width=1cm,align=center,text=black!75] {ethernet};

\draw [-Stealth] (fpga.east) -- (card.west);
\begin{pgfonlayer}{background}
	\draw[black!50, line width=3pt] (-2,3) -- ++ (11,0) node[below left]  {high voltage, magnetic field}  -- ++ (0,-6) -- ++(-11,0);
	\draw[black!50, line width=3pt] (-2,2.5) -- ++ (5,0) node[below left]  {high vacuum}  -- ++ (0,-5) -- ++(-5,0);
\end{pgfonlayer} 
\end{tikzpicture}
}
\caption{Operational structure of an RADC DAQ system in the KATRIN detector section. Environmental constraints from the high voltage and magnetic field are outlined in gray and isolated to the front-end board, which provides detector bias voltages and digitizes the detector signals with minimized analog paths. Connection via optical links to a back-end unit allows for flexible event filtering and readout mode implementation.}
\label{fig:RADC_concept}
\end{figure}

The minimization of analog signal paths is desired to decrease the coupling of electronics noise into the detector waveforms, which decreases the energy resolution of the TRISTAN detector.
Therefore, the front-end board is mounted directly on the vacuum chamber and is designed to be operated in a strong  magnetic field and insulated from  high-voltage. 
The front-end board digitizes and packetizes the detector waveforms, which are sent via high-speed fiber optic data links to the back-end FPGA boards, which are located outside of the stray magnetic fields of the detector magnet and on earth ground potential. 

Waveform signals are then processed through digital signal processing architectures and event creation logic implemented on the FPGA and computer server.
Digital signal filtering creates events from detector pulses with associated timing and energy evaluations, charge sharing and pileup flagging, and other pulse attributes.
Data from the back-end are processed through a high-performance Ethernet-based DAQ system that interfaces with the computer server and data storage.

The separation of the two stages allows for the development of a project-specific front-end board that meet the operational design requirements of integrating the TRISTAN detector into the KATRIN detector section. 
Removing these design requirements from the back-end allows the use of more standardized FPGA boards without a large additional development.

\section{Front-End DAQ System}\label{sec:frontend}
The primary component of the front-end DAQ system is the Tile Main Board (TMB); a project-specific board designed by the Karlsruhe Institute of Technology (KIT) Institute for Data Processing and Electronics (IPE).
Each TMB is designed to digitize and transmit detector signals and provide slow control for a single 166-pixel detector module. 
The TMB is connected onto the detector vacuum chamber by four 100-pin micro-D connectors, providing connections for detector signal and sensor readout as well as the in-vacuum control voltages. 
The location at the detector chamber sits on a high voltage of up to $\SI{25}{\kilo\volt}$, necessary for the operation of a post-acceleration electrode (PAE), and is exposed to an $\mathcal{O}(\SI{100}{\milli\tesla})$ stray magnetic field of the detector magnet \cite{Amsbaugh_2015}. 
Therefore, the TMB excludes any ferromagnetic components and is powered with isolation transformers. 
All data transfer between a TMB and the back-end FPGA board is done via fiber optic links.

\begin{figure}[ht!]
    \centering
               \includegraphics[width=.9\textwidth]{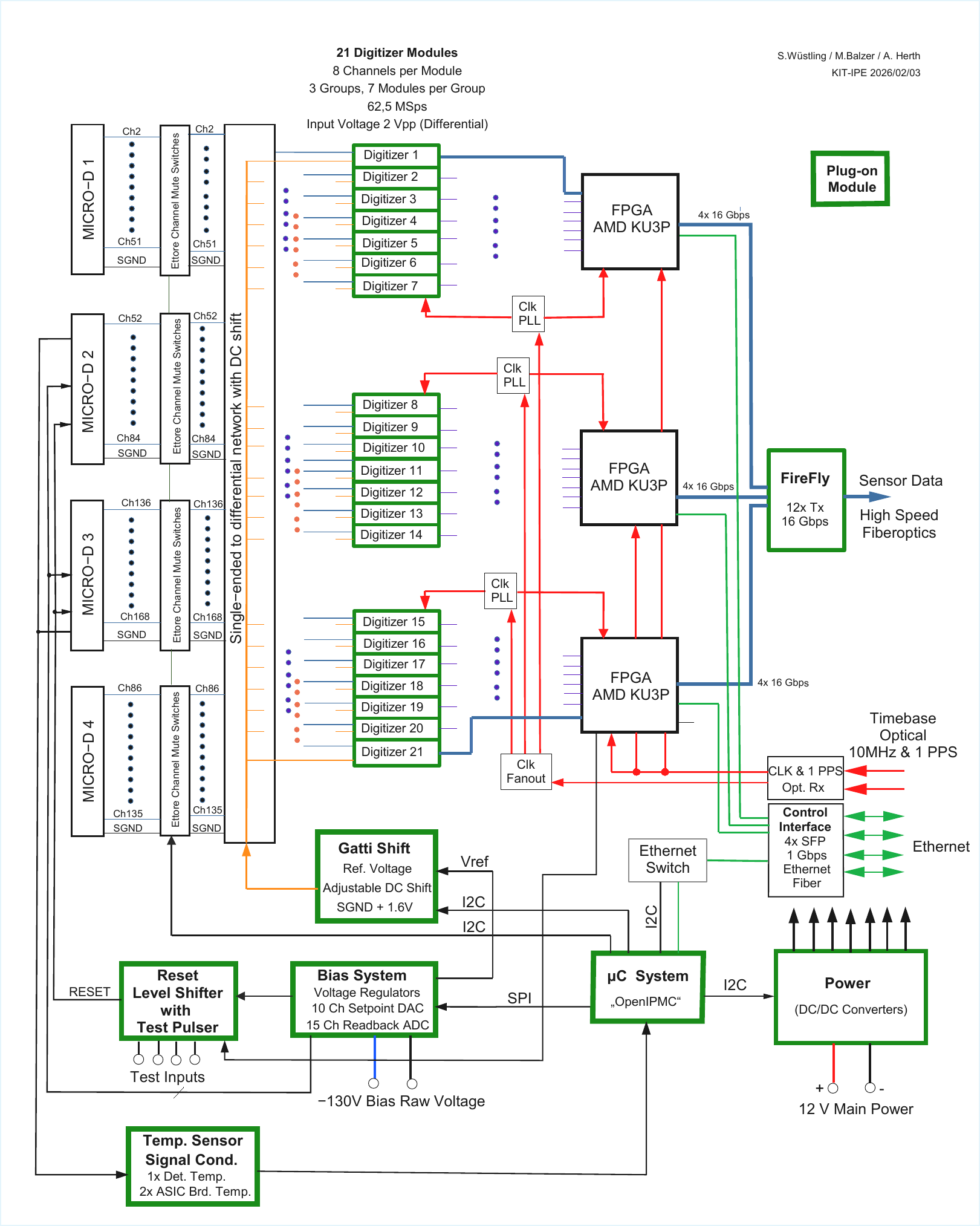}
    \caption{Block diagram of the Tile Main Board (TMB). Green blocks indicate individual plug-on submodules which populate the PCB. The TMB provides detector operating voltage and waveform readout through the four micro-D connectors and streams waveform data to the back-end through fiber optic transmitters.}
    \label{fig:TMB_schematic}
\end{figure}

The TMB consists of a $\SI{411}{\milli\meter} \times \SI{340}{\milli\meter}$ PCB that is populated with submodules, each with a specific readout or control task.
The separation of tasks into smaller submodules allows for easier development, troubleshooting, and repair. 
The TMB submodules include: Microcontroller ($\mu$C), Power Supply, Bias, RESET, voltage offset or Gatti Slider, and ADC Boards.  
All modules are custom designed for this project except for the $\mu$C board.
The components and connections of the TMB's submodules are shown in \autoref{fig:TMB_schematic}, with an example TMB installed on a detector chamber shown in \autoref{fig:TMB_replica}.
The structure and role of each of the submodules are given below.

\paragraph{$\mu$C Board}
The TMB operation and slow control readout is performed over a $\SI{1}{\giga\bit\per\second}$ QSFP ethernet connection.
The board control is based on the open-source OpenIPMC software, which allows full user customization of board control and debugging \cite{9465210}.
A $\mu$C mezzanine, tested on ATCA boards designed for projects at the HiLumi-LHC, distributes the slow control parameters to their respective submodules, the Bias Supply, Gatti Slider Boards, and Power Supply Board \cite{Calligaris_2022}.
The readout of temperature sensors on the detector and ASIC are also processed through the $\mu$C Board.

\paragraph{Power Supply Board}
All components of the TMB are powered through a Power Supply Board that is fed with a supply voltage of $\SI{12}{\volt}$ and a supply current of $\SI{15}{\ampere}$. 
Air-coil DC/DC converters are used to derive the necessary voltages for operating each of the TMB submodules and FPGAs. 

\paragraph{Bias Board}
The bias and supply voltages necessary for operating the SDDs are communicated from the $\mu$C Board to the Bias Board. 
The Bias Board consists of 10 digital to analog converters, which supply the necessary voltages to the middle two micro-D vacuum feedthroughs. 
A 15 channel ADC is also implemented for real time slow control readback of the detector operating voltages. 

\paragraph{Gatti Slider Board}
A reference voltage for applying a DC-offset to the detector signals before digitization is provided by the Gatti Slider Board.
The offset is necessary to map the voltage output of the detector into the operating range of the ADC boards.
The value of the offset is controlled by the slow control parameters of readout operation and is changed between measurements to randomize the pulse positions within the digitization range \cite{Cottini1963ANM,Gatti_1969}. 
The addition of this board was necessitated by investigations of the impact of ADC nonlinearities on event energy reconstruction \cite{Dolde_2017,Abgrall_2021}. 
The Gatti Slider Board effectively introduces a sliding scale linearization of the readout over an extended measurement time.
A voltage stability of $< \SI{0.5}{\milli\volt}$ is necessary over the digitization of individual detector pulses to not introduce substantial bias to the energy reconstruction.

\paragraph{RESET Board}

In order to avoid saturation from accumulating charge, the detector is periodically reset at a fixed frequency by the RESET Board.
The signal readout is inhibited in the ETTORE ASICs and the detector anodes are discharged to return to the normal detector operating point. 
The RESET board also allows for injection of test signals from an outside signal source to the readout chain. 

\begin{figure}[t!]
    \centering
    \includegraphics[width=.8\textwidth]{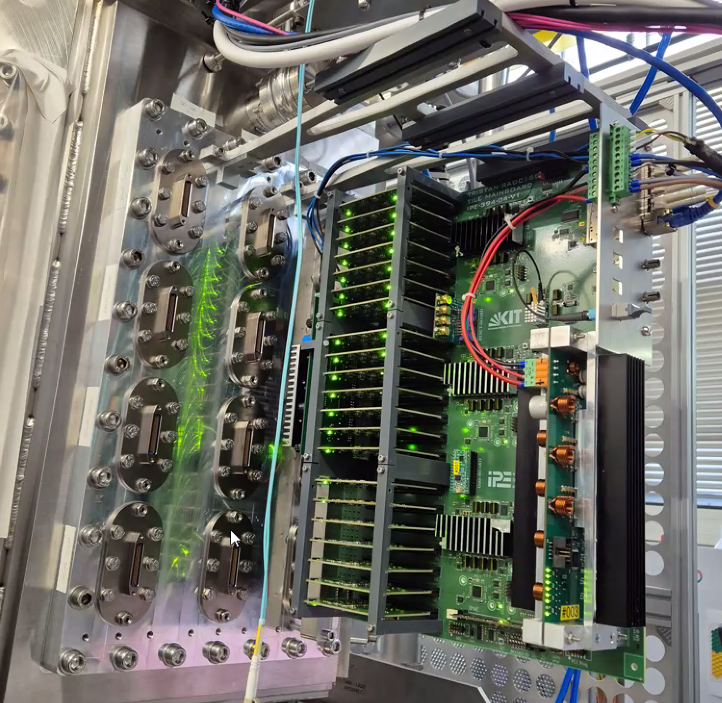}
    \caption{Tile Main Board (TMB) installed and operated at a replica of the detector chamber upgrade. 21 ADC cards are mounted perpendicular to the TMB to allow for dense PCB design. Fiber optic connections on the right side provide connections to back-end control and readout. Two mounting positions for additional TMBs are visible to the left; each a column of four micro-D vacuum feedthroughs.}
    \label{fig:TMB_replica}
\end{figure}

\paragraph{ADC Boards}
Readout of the analog signals from a 166-pixel detector is performed by 21 ADC Boards and three FPGAs. 
Slow control parameters for the sampling frequency and downsampling ratio are set in the FPGAs via the SFP module's $\SI{1}{\giga\bit\per\second}$ ethernet connection. 
Each ADC Board is composed of an anti-aliasing filter and an 8-channel AD9257 ADC.\footnote{https://www.analog.com/en/products/ad9257.html} 
A 5th order Bessel low-pass filter is used due to it being well suited for pulse processing \cite{1999AN6A}. 
The AD9257 provides 14-bit sampling of the detector signals in a $\SI{2}{\volt}_{\text{pp}}$ range at a sampling frequency of up to $\SI{65}{\mega\hertz}$. 
Each ADC Board streams data at a rate of $\SI{6.78}{\giga\bit\per\second}$ to the FPGA.

\paragraph{FPGAs and Transmitters}
Each TMB is equipped with three FPGAs, which control the processing and transfer of detector waveforms.
An additional bit of information is added in the FPGA per sample indicating whether an inhibit signal is present, either from the RESET Board or from an external input. 
The front end FPGAs serialize and format the data from the ADC Boards for fiber optic transmission. 
The streaming of data from the FPGAs to the back-end is performed by 12 Samtec FireFly™ ECUO optical transmitters.\footnote{https://suddendocs.samtec.com/literature/samtec\_high\_speed\_cable\_guide.pdf}
Each transmitter operates at up to $\SI{16}{\giga\bit\per\second}$, providing an overall streaming rate of $\SI{192}{\giga\bit\per\second}$ in order to directly stream all 168 ADC channels.

\paragraph{Clock System}
Time synchronization between components within one TMB, multiple TMBs, and the back-end system is controlled by a master GPS clock. 
The GPS clock provides a fanout of $1$ Pulse Per Second (PPS) and $\SI{10}{\mega\hertz}$ signals via fiber optic lines to the DAQ system. 
Within a single TMB, the timing signals are distributed through a clock fanout and FPGAs and their respective ADC cards are connected with a phase-locked loop.
Synchronization between the ADCs and DC/DC converters are also important to avoid forming beat frequencies.
The 1 PPS signal is embedded within the TMB data stream, which allows the back-end systems to align their reproduced event timestamps while running asynchronously.

\section{Back-End DAQ System}\label{sec:backend}
Transmission via fiber optic cables allows the back-end system to be located outside of the stray magnetic fields and off the post-acceleration high voltage of the detector chamber, lifting the environmental design requirements needed for the front-end board from the back-end system.
The function of the back-end system is to provide multiple readout modes of TRISTAN detector data through digital signal processing and event creation.

The hardware platform for the back-end system is the Serenity-S1 FPGA ATCA-card developed by KIT-IPE and CMS collaborators for the CMS detector at the LHC \cite{Rose:2019oiy,Mehner_2024}. 
Serenity-S1 cards are stored in ATCA racks and provide the detector signal processing through an AMD Virtex UltraScale+ FPGA.\footnote{https://docs.amd.com/r/en-US/ds923-virtex-ultrascale-plus/}
The VU9P version is used in the context of the TRISTAN DAQ.
Each Serenity-S1 card is capable of processing the data from three TMBs, providing the readout of 498 TRISTAN detector pixels.
The readout volume per Serenity-S1  is limited primarily by the routing resources of the FPGA.
Digital signal processing on the FPGA allows for the application of tailored filtering algorithms on the transmitted waveforms.

Four readout modes; Waveform, ListWave, ListMode, and Histogram, are implemented in the back-end processing. 
Waveform readout directly saves all data streamed from the front-end. 
ListWave and ListMode readouts record information from triggered detector events with and without an associated waveform snippet, respectively. 
Histogram readout populates predefined energy histograms with the triggered events based on decision tree logic. 
While science data taking will primarily be taken using the Histogram readout, the other readout modes enable systematic studies of detector and data processing related effects.

The event filtering logic presented in \autoref{sec:filters} is implemented on the back-end FPGA. 
Filter parameters are adjustable to the user through setting of fixed FPGA registers. 
The parameters of the trigger and energy filtering are independently controllable for each detector readout channel, which allows for optimization to account for the nonhomogeneous detector characteristics.
For event triggering, $\tau_w$ can be set within a range of  $\SI{64}{\nano\second}$ to $\SI{1}{\micro\second}$ with any required threshold value.
The energy filter $\tau_p$ and $\tau_g$ are individually adjustable within the ranges of $\SI{64}{\nano\second}$ to $\SI{8}{\micro\second}$, and $\SI{0}{\nano\second}$ to $\SI{1024}{\nano\second}$, respectively. 
The sharing of trigger information between readout channels is performed through user implemented trigger distribution maps for channels within the same Serenity-S1 card.

An event builder takes triggers  and produces event information packets which are saved in ListWave and ListMode readout modes.
Each event packet consists of a header with 32 bytes of information: an 8 byte timestamp, 4 byte energy evaluation, 2 byte channel number, 2 byte flag, and 16 bytes of additional waveform information.
The additional 16 bytes are reserved for waveform parameters that the user can configure in the readout. 
In ListWave mode the length of the saved waveform associated to the event is adjustable up to $\mathcal{O}(\SI{10}{\micro\second})$ in order to fully sample the rising edge and decaying tail of the detector waveform. 

In Histogram readout mode the events from ListMode are used to populate histograms corresponding to their energy evaluation, channel number, and flags.
Production of multiple histograms from the same data, incorporating different flag combinations, allows for the systematic study of different effects on the measured energy spectrum.
The dynamic range of the energy histograms, as well as the energy resolution of each histogram bin, is set by the user.
The analysis of the full tritium spectrum will utilize data produced by an optimized set of histogram parameters and flags to minimize systematic effects.

Each Serenity-S1 card is connected to the computer server via a \SI{100}{\giga\bit\per\second} uplink.
A challenging part of the readout is handling the streaming of the possible hundreds of $\SI{}{\giga\bit\per\second}$ from the Serenity-S1 cards with zero data loss.
To overcome this challenge, the Data Acquisition Development Kit (DQDK), a software framework for high-performance Ethernet-based data acquisition, was developed.
The Serenity-S1 card runs a resource-efficient implementation of the Ethernet data transfer protocols: User Datagram Protocol (UDP), Internet Protocol version 4 (IPv4) over MAC \& PMA/PCS Xilinx IP core.
UDP was chosen due to its FPGA resource efficiency when compared to the alternative Transmission Control Protocol.
The implementation utilizes only 1,048 of the available 67,200 configurable logic blocks and zero Block RAM resources \cite{10659873}.
Possible losses due to congestion on the Network Interface Card or computer are avoided by using AF\_XDP sockets to ensure fast receiving of data. 
In addition, software optimizations and computer architectural support enhance performance through parallelism and overcome possible memory bottlenecks \cite{10329590}.
The combination of these methods has been empirically shown to prevent data loss.
 
In Waveform, ListWave, and ListMode operation the data are copied into a larger intermediate buffer before being directly saved given the run length specifications of the user. 
In Histogram mode the streaming of data from the Serenity-S1 card to computer server is identical to ListMode, however, instead of being saved into a buffer, the incoming events are accumulated into their respective histograms.
The operation of all readout modes require $\SI{128}{\giga\byte}$ of RAM on the server for each TRISTAN module.

The system is designed with sufficient overhead in order to safely handle the high rate of data streaming and processing. 
The utilization rates of the FPGA RAM and internal logic resources, such as look-up tables and flip-flops, remains below $50\%$, and below $33\%$ for the Serenity-S1 transceiver limit.
The highest utilization rate of approximately $75\%$ is in the digital signal processing from the constant stream of waveform data from the front-end to the Serenity-S1.

\section{TRISTAN Operation}

The projected sensitivity for the TRISTAN detector upgrade is based on a four month measurement of the tritium $\upbeta$-decay spectrum \cite{acharya2026katrinsensitivitykevsterile}. 
Operation of the Waveform readout produces a data rate of $\SI{125}{\mega\byte\per\second}$ per pixel, which results in an overall data rate of $\SI{186.7}{\giga\byte\per\second}$ for nine modules.
A reduction of data is evidently necessary to be able to manage and analyze the measurement.
Recording the events of $10^5$ incident electrons per second per pixel in ListMode readout reduces the overall data rate to  $\SI{4.78}{\giga\byte\per\second}$ but for a measurement campaign of four months this corresponds to $\mathcal{O}(\SI{10}{\peta\byte})$ of saved data.
A large reduction in the size of the data produced comes from operating in Histogram mode but requires the loss of information on the event level as well as on the temporal level shorter than the accumulation time of each histogram.

The planned histograms for the TRISTAN case will consist of $2^{16}$ bins spanning a dynamic range of $\SI{200}{\kilo\electronvolt}$. 
A user configurable bin-depth of 2-, 4-, or 8-bytes can be set depending on the measurement configuration and the accumulation time. 
Simulation of the TRISTAN detector response in its normal operating conditions indicate that for a 2-byte histogram an overflow is expected after $2004\pm 8$ seconds of operation \cite{Gavin_dissertation}. 
In the event of an overflow of the bin-depth, the bin reaches a saturation point and additional counts are not incremented. 
Events above the maximum energy of the histogram are assumed to be rare and counted in the last energy bin. 

\begin{figure}[ht!]
\includegraphics[width=\columnwidth]{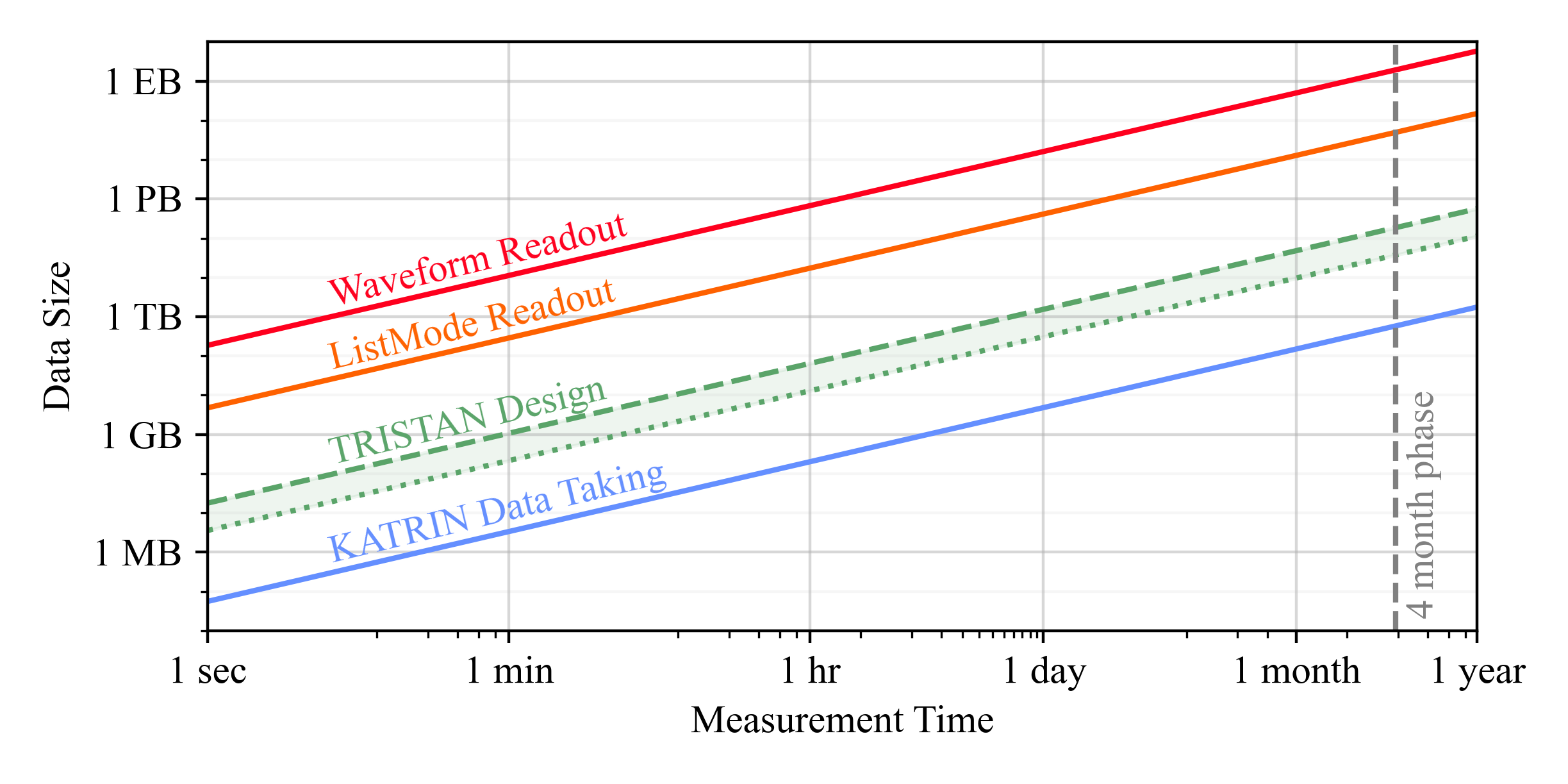}
\caption{Accumulated total data size as a function of measurement time for different data taking readout modes. The designed TRISTAN readout structure is shown to reduce the overall data size by three and five orders of magnitude compared to the ListMode and Waveform readouts, respectively. The range TRISTAN rates represents the difference between uncompressed and compressed data saving.}
\label{fig:datarates}
\end{figure}

The data volume produced as a function of measurement time for the various operating modes, compared to the data rate of KATRIN neutrino mass measurement, is shown in \autoref{fig:datarates}.
A measurement distribution of the three readout modes will be implemented for the TRISTAN measurement to keep overall data rates manageable while retaining information used for performance and stability checks. 
The measurement time is divided into roughly one hour runs.
The majority of the time consists of the accumulation of twelve histograms, each with five minute accumulation times to avoid possible histogram overflows.
The timing resolution of the histograms over the run allows for temporal resolution of systematic effects.
Each run also has an accompanying one minute and ten millisecond ListMode and Waveform readout, respectively.
Application of this data taking strategy reduces the saved data of the four month measurement to $\mathcal{O}(\SI{100}{\tera\byte})$.
Further reduction of the data size can be achieved if fewer histograms or longer histogram accumulation times are used, or through the application of compression algorithms on the sparse energy histograms.

\section{Conclusion}

To meet the experimental challenges of operating the TRISTAN detector upgrade to the KATRIN experiment a custom RADC DAQ system is developed that physically separates signal digitization and processing components.
The magnetic field and high-voltage compatible front-end TMB is designed to operate the TRISTAN detectors while reading back slow control parameters and performing waveform digitization.
The Gatti Slider Board of the TMB allows for analog offsets before digitization to reduce the impact of nonlinearities in the reconstructed event energies.

The back-end of the DAQ, composed of Serenity-S1 FGPA cards and a computer server, allows for flexible digital pulse processing.
Application of a custom trigger logic, trigger map, and event flagging system allows for accurate energy evaluations and mitigation of event pileup and inter-pixel effects.
The multiple operational modes of the RADC DAQ system provide a scalable readout of the $\mathcal{O}(1000)$ channels with reasonable output data rates for post-processing and analysis.
 
\acknowledgments 

This material is based upon work supported by the U.S. Department of Energy, Office of Science, Office of Nuclear Physics under Award Numbers DE-FG02-97ER41041, and DE-FG02-97ER41033, and by the National Science Foundation under Grant No. NSF OISE 1743790.
This project has received funding from the European Research Council (ERC) under the European Union Horizon 2020 research and innovation programme (grant agreement no. 852845).
We acknowledge that we made use of the Serenity-S1 cards developed by the Serenity Collaboration for the high-luminosity upgrade of the CMS experiment at the LHC. 

\bibliographystyle{JHEP}
\bibliography{sources.bib}

\end{document}